\documentclass[fleqn,10pt]{olplainarticle}
\usepackage{placeins}
\usepackage{url}

\title{Corrigendum and addendum to: How Populist are Parties? Measuring Degrees of Populism in Party Manifestos Using Supervised Machine Learning}

\author[1]{Jessica Di Cocco}
\author[2]{Bernardo Monechi}
\affil[1]{European University Institute, Via della Badia dei Roccettini, 9, 50014 Fiesole (FI). E-mail: jessica.dicocco@eui.eu}
\affil[2]{Sony Computer Science Laboratories, 6 Rue Amyot, 75005 Paris, France}


\begin{abstract}
This paper is a corrigendum and addendum to the previously published article: 'How Populist are Parties? Measuring Degrees of Populism in Party Manifestos Using Supervised Machine Learning' (\textit{Political Analysis}, 1-17. doi:10.1017/pan.2021.29). These corrigendum and addendum were prepared to correct errors in data labelling and show some extra insights not included in the previously published paper. Here, we report these corrections and point to some additional conclusions by focusing on the effects of the label reshuffling per parties and years and presenting new figures wherever appropriate. We show that although the simplified labelling method proposed in the previously-published article can induce biases in the correlations with expert scores, random labelling reduces correlations significantly. We show that this is also true for correlations based on a manually-coded data set.  
These modifications are based on other evidence and results reported in detail in a future publication.

\end{abstract}

\begin{document}

\flushbottom
\maketitle
\thispagestyle{empty}

\section*{Introduction}
One of the main challenges in comparative studies on populism concerns its temporal and spatial measurements within and between many parties and countries. Textual analysis has proved helpful for these purposes, and automated methods can further improve research in this direction. In the paper published in \textit{Political Analysis}, we proposed a method to derive a score of parties' levels of populism \citep{di2021populist}. We used supervised machine learning to perform textual analysis on national manifestos and derive the score, which measures the distance between manifestos of populist and non-populist parties. For the score derivation, we used the Random Forest classification algorithm \citep{breiman2001random}. The final party score is the fraction of its manifesto's sentences that the classifier considers as belonging to a prototypically populist party manifesto in its nation.
The advantages of using text-as-data in the measurement of populism are several (for a detailed discussion on this, see \citet{hawkins2019measuring}). For example, they allow focusing on the elites and their ideas; inferring political actors' position directly from their texts; measuring populism across a large number of cases, within and between countries; and obtaining continuous populism measures which, unlike dichotomous ones, better account for the multi-dimensionality of populism and differentiate between its degrees \citep{meijers2021measuring}. 

Without a monolingual corpus, we performed separate training for each country, obtaining six different models. For data labelling, we gave the label $Y=1$ to all the sentences belonging to manifestos of parties recognized as populist in the PopuList classification \citep{rooduijn2019populist}. We also gave the label $Y=1$ to all the sentences belonging to Power to People (PaP) in Italy since it exhibits some populist features in a country where all the pre-existing populist parties lean toward other ideological alignments. At this stage, we excluded from the data the manifestos of those parties that are ambiguously populist over time or are considered as populist only in some classifications.
With the labelled text data, we built models capable of assigning its corresponding label to each chunk of text \citep{alpaydin2020introduction}. Considering the training set of a country, we performed a "Grid Search" over a set of hyperparameters of the Random Forest algorithm to find the best combination according to a classification accuracy metric. In other words, we iterated all the combinations of the chosen hyperparameters, selecting the most accurate one. Once we found the best hyperparameters' combination, we retrained the model on the whole training set. Finally, we used the entire training set to find the best threshold for the probabilities given by the Random Forest algorithm using the Receiver Operating Characteristic Curve (ROC) and the Youden index \citep{ruopp2008youden}. This latter procedure further increases the model's final accuracy, choosing the best combination of true positive and false positive rates.
When validating this method, we illustrated that the score is strongly correlated with key attributes of populism (e.g., anti-elitism and people centrism) drawn from the 2017 Chapel Hill Expert Survey (CHES) \cite{doi:10.1177/2053168016686915} and the 2018 Populism and Political Parties Expert Survey (POPPA) \citep{meijers2020measuring}. We also validated the score against the latent populism variable in the POPPA data, constituted by five components of populism identified by the ideational approach (anti-elitism, people-centrism, Manichean worldview, general will, and indivisible people) as suggested by Meijers and Zaslove \citet{meijers2020populism}.

This corrigendum and addendum were prepared to correct some errors in data labelling and show some extra insights not included in the previously published paper. Errors in the data labelling referred to the Austrian case. The corrected and updated database can be found here~\citep{new_dataset}. Here, we also point to some additional conclusions by focusing on the effects of the label reshuffling per parties and years. We show that although the simplified labelling method proposed in our article \citep{di2021populist} can induce biases in the correlations with expert scores, random labelling reduces correlations significantly. We show that this is also true for correlations based on a manually-coded data set.  

\section*{Results}

\subsection*{Corrigendum}
The dataset uploaded on the Political Analysis Dataverse \citep{DVN/BMJYAN_2021} had some labelling errors that, however, did not alter the results from what is shown in \citet{di2021populist}. The labelling error concerned mainly the Austrian Manifestos. The party-year labels were randomly reassigned to the manifestos so that 'populist' manifestos \footnote{We refer to manifestos by parties usually considered as populists. For further details and a deeper discussion about this, see \citet{di2021populist}} could have been assigned to non-populist parties. This reshuffled data were part of unpublished results that we decided not to include in our paper. Moreover, there was a repeated manifesto for the German data due to two parties being in a coalition during the same electoral round. We would like to thank Michael Jankowski (Institute for Social Sciences, University of Bremen) and Robert A. Huber (Department of Political Science, University of Salzburg) for informing us about these mistakes. 
To prevent other researchers from using the wrong results for their work, we decided to upload the correct version of the data in a new repository on GitHub~ \citep{new_dataset}, together with the correct scores and the new results shown in the following part of this addendum. 
Note that to reduce training time, the new scores have been computed using the Gradient Boosting Algorithm \citep{friedman2001greedy} instead of Random Forest Algorithm \citep{breiman2001random}. We have shown in \citet{di2021populist} that the results are pretty independent of the chosen algorithm. However, short training time is mandatory to study the effect of label reshuffling that needs hundreds of models to be trained. In Table~\ref{tab:old_valid_check} we show the correlations between our score, the 'anti-elite salience' and 'people vs elite' dimensions of 2017 CHES \citep{doi:10.1177/2053168016686915}, and a latent populism variable constructed using five dimensions of the 2018 POPPA data as suggested by \citep{meijers2020measuring}. We used either the score computed on the correct data and the one with label reshuffling for the Austrian case.
\begin{table}[ht]
\centering
\begin{tabular}{|c|c|c|}
\hline
 & With Reshuffled AT  & Correct Data \\
\hline
CHES ‘anti-elite salience’ & $0.77$ with CI$\;=[0.60, 0.87]$ & $0.816$ with CI$\;=[0.669,0.902]$\\
\hline
CHES ‘people vs elite’ & $0.81$ with CI$\;=[0.67,0.90]$ &   $0.715$ with CI$\;=[0.509,0.844]$\\
\hline
POPPA latent variable & $0.85$ with CI$\;=[0.71, 0.92]$ & $0.818$ with CI$\;=[0.683,0.9]$ \\
\hline
\end{tabular}
\caption{\label{tab:old_valid_check} Pearson's coefficient between the party Pop. Score introduced 'anti-elite salience' and 'people vs elite' dimensions of 2017 CHES \citep{doi:10.1177/2053168016686915}, and the latent populism variable computed using 2018 POPPA \citep{meijers2020measuring}. The $95\%$ confidence interval is also shown in each cell. In the first column, we used the scores computed in \citet{di2021populist} with the reshuffled label for Austrian manifestos. In the second one, we used the scores computed with the correct data, changing the classification algorithm from Random Forest to Gradient Boosting. Moreover, we removed numbers from the sentences.}
\end{table}
We can see that the correlations are significant and compatible in both cases. Austrian parties are not part of the CHES data, so in this case, manifesto relabelling cannot affect the results, and the observed variations are mainly due to the change of the classification algorithms. Concerning the POPPA latent variable, the agreement between the results is partly due to the reshuffling of only the Austrian data (6 parties out of 41), but also to a bias that our method might induce in the party score. Thus, we could argue if our method always produces correlations with experts' scores independently from the content of the manifestos.

\subsection*{Addendum}

We focus on the effects of the label reshuffling per parties and years to investigate if correlations between the score of populism and the expert surveys persist also when texts are randomly labelled. The rationale for this investigation is that it is always possible to divide texts in two groups and train a classifier able to discriminate whether a sentence belongs to a group or another. This aspect can result into a bias involving the score, which could be correlated with validation datasets (i.e. expert surveys in our case) even when text labelling is random. Hence, sentences that we assigned to populist parties will always have higher probability to get a high score of populism. On the contrary, sentences taken from texts of non-populist parties will always have a higher probability to obtain a high score of populism. In our article, we highlighted the need for a manually coded dataset to obtain more accurate results. Nevertheless, performing manual coding is resource intensive and requires several content and language skills when dealing with multilingual texts and populism. After looking at the Italian case, for which we also had a manually coded dataset, we concluded that our simplified method of labelling can still prove useful for our purposes. However, this method might lead to spurious correlations when it is validated with external scores, such as the expert surveys, even in the case of incorrect manifesto labelling. 
To investigate this issue, we use 100 samples of manifestos with reshuffled labels. When reshuffling the labels, we randomly exchange the \emph{party-year} labels between all the manifestos of the same country. In this way, we can generate samples were ``populist'' manifestos might actually have no populist content, while populist content could be shared within ``non-populist'' manifestos.
%
For each reshuffled sample, we repeat the exact procedure from~\cite{di2021populist} to derive a score for each party in each year. 
We suppose that correlations observed when looking at reshuffled sample score and expert surveys should be less significant than those obtained with the correct labelling. We test this hypothesis using the 2017 CHES \citep{doi:10.1177/2053168016686915} and the 2018 POPPA \citep{meijers2020populism}. When looking at the POPPA, we take into consideration the latent populism variable constructed using five dimensions as suggested by \citet{meijers2021measuring}. These dimensions include the Manichean vision of politics, the indivisibility of the ordinary people, people’s general will, people-centrism, and anti-elitism. 
Following the results in \citet{di2021populist}, we also compared the reshuffled scores to another score obtained using a manually coded dataset for the Italian case. In \citet{di2021populist} we showed a good agreement between our score and the manually coded one. Again, our hypothesis is that such correlations are less strong in the reshuffled case indicating that manifesto content is relevant to derive a correct party score.
\subsubsection*{Expert Score Correlations}

\begin{figure}[ht]
\centering
\includegraphics[width=0.9\linewidth]{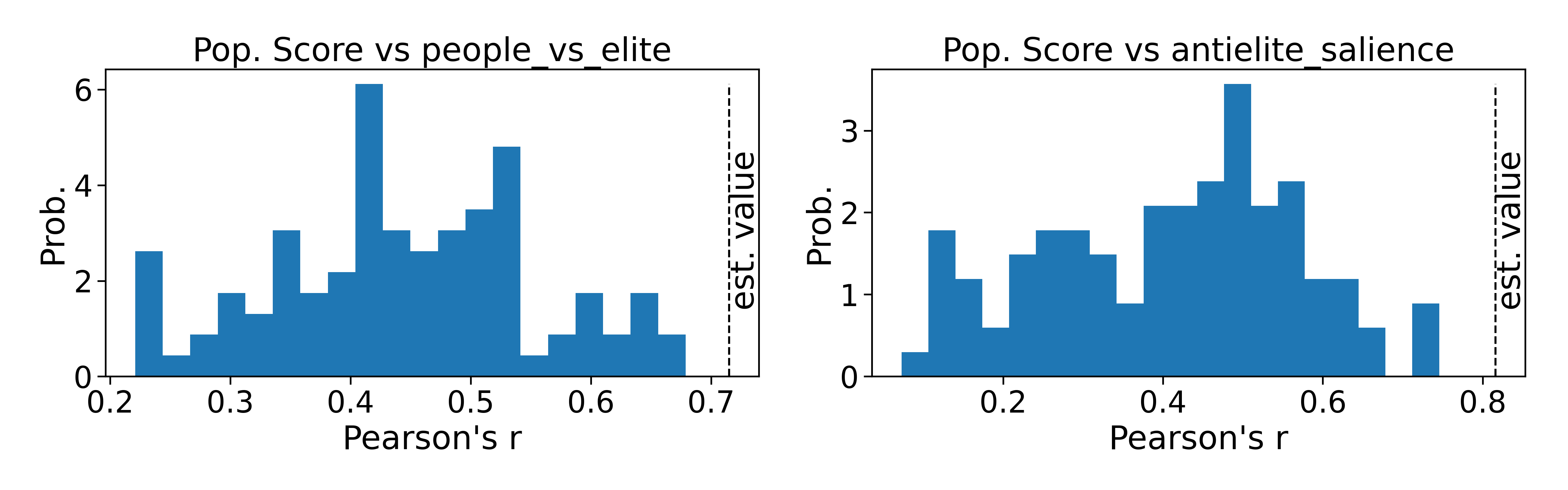}
\caption{Pearson's correlation coefficient between the CHES ‘people vs elite’(left) or ‘anti-elite salience’ (right) and the Populist Score computed using the 100 reshuffled manifesto data. The vertical dashed line is the coefficient value for the non-reshuffled dataset also shown in Table~\ref{tab:old_valid_check}.}
\label{fig:ches_resh}
\end{figure}

Fig.~\ref{fig:ches_resh} shows the distributions of the Pearson's correlation coefficients between the score and the two CHES selected dimensions \citep{doi:10.1177/2053168016686915}, obtained from the 100 reshuffled datasets. In each panel, we also show the estimated value obtained with the correct data and shown in Table~\ref{tab:old_valid_check}. Some positive correlations can still be found using the reshuffled data, they are always considerably lower than those obtained with the non-reshuffled case. This indicates that, despite our method could induces a bias in the party score, the result is not independent from the manifesto content.

\begin{figure}[ht]
\centering
\includegraphics[width=0.4\linewidth]{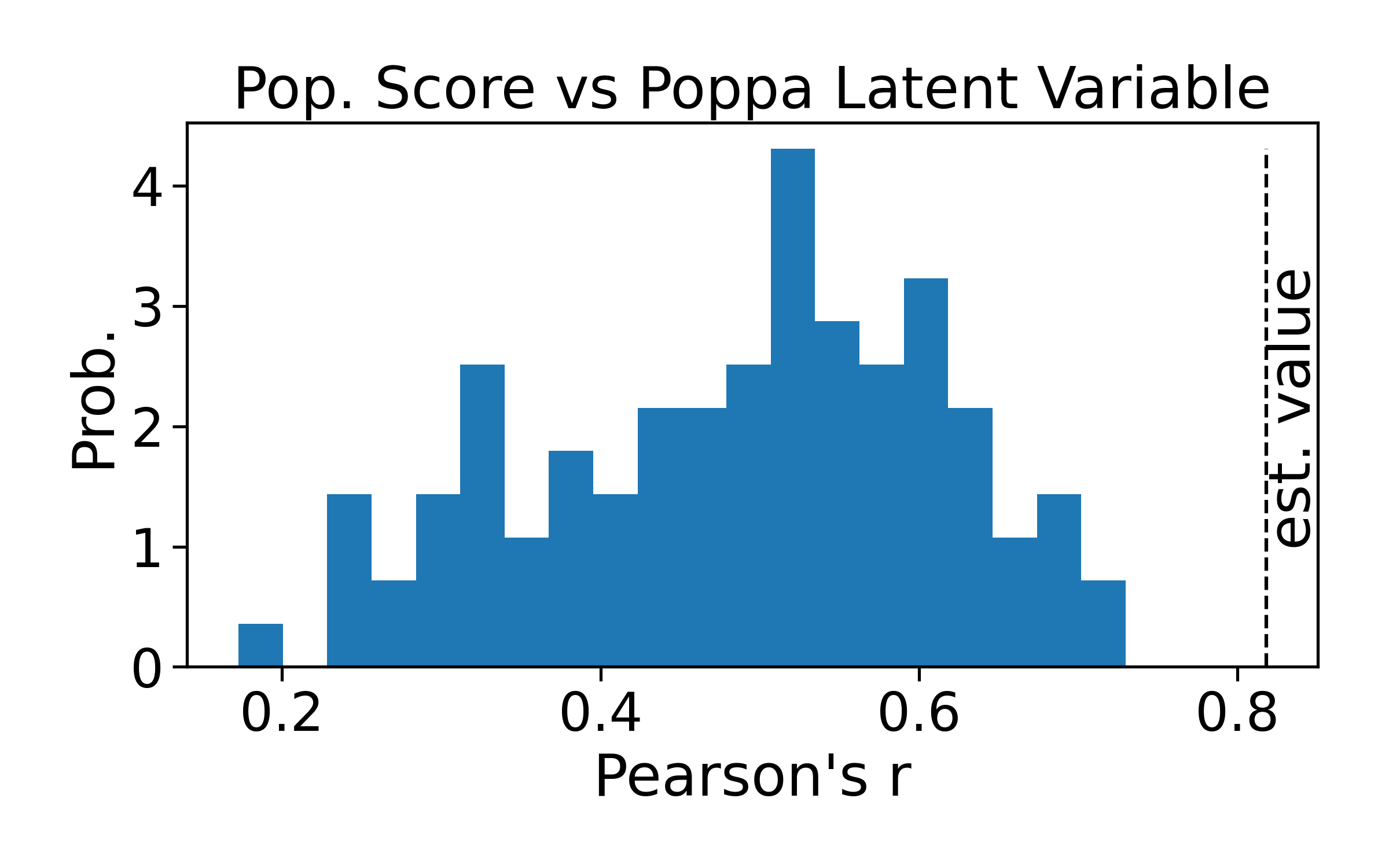}
\caption{Pearson's correlation coefficient between the POPPA latent variable and the Populist Score computed using the 100 reshuffled manifesto data. The vertical dashed line is the coefficient value for the non-reshuffled dataset also shown in Table~\ref{tab:old_valid_check}.}
\label{fig:poppa_resh}
\end{figure}

Similar results can be found in Fig.~\ref{fig:poppa_resh}, where we used the 2018 POPPA latent variable \citep{meijers2020measuring}.
\FloatBarrier
\subsubsection*{Manual Coding Score Correlations}
\begin{figure}[ht]
\centering
\includegraphics[width=0.45\linewidth]{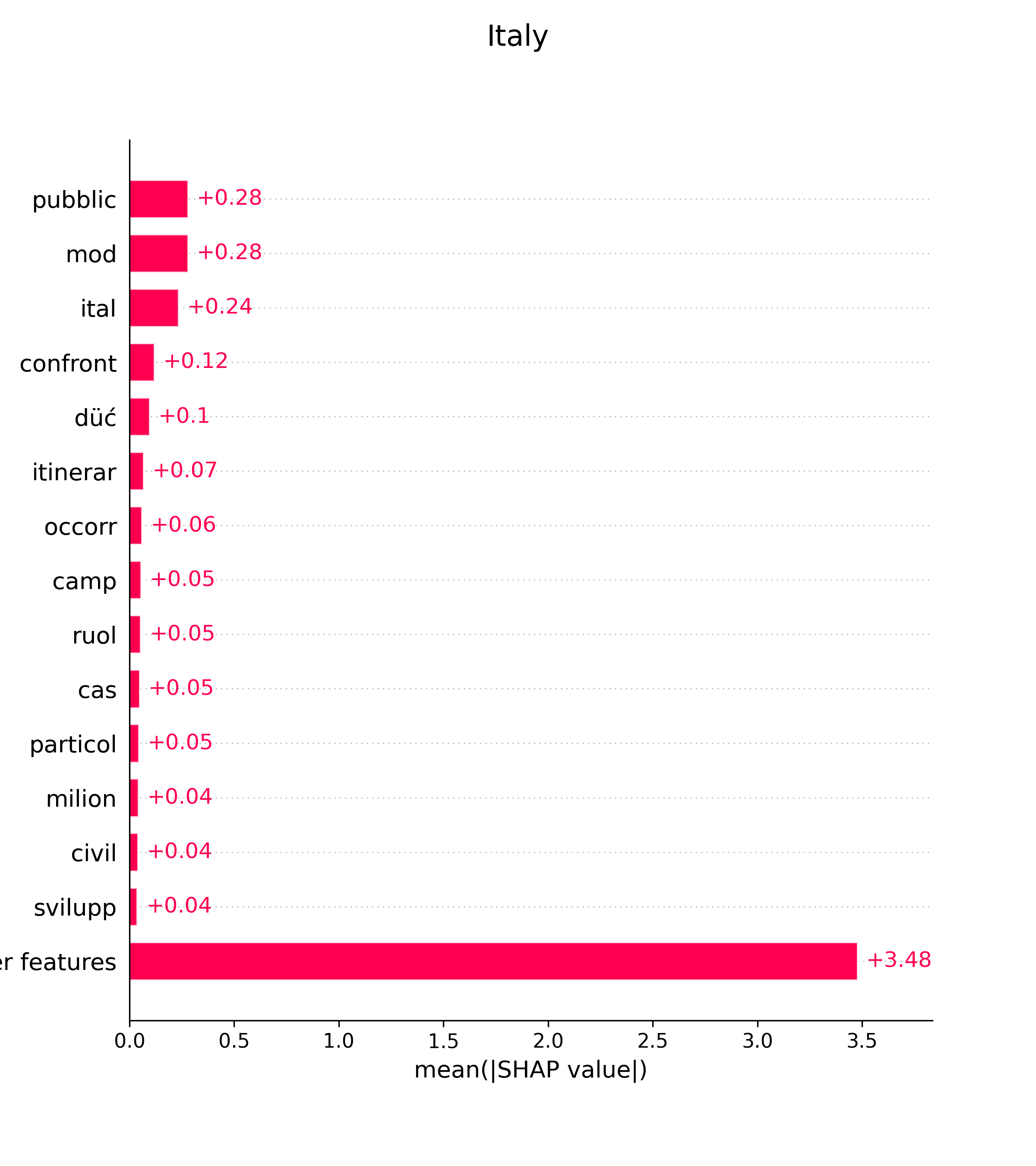}
\includegraphics[width=0.45\linewidth]{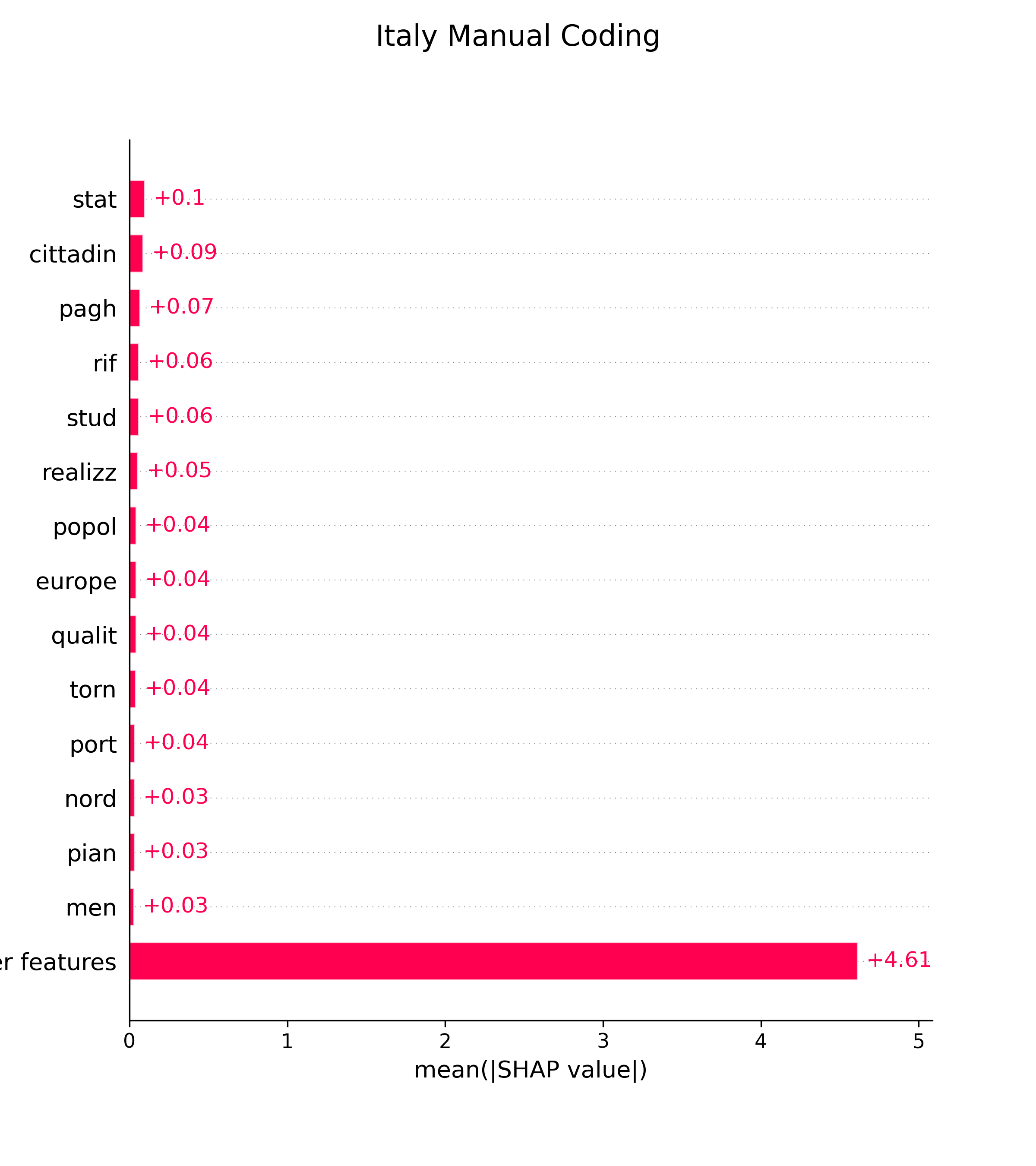}
\caption{The 15 most relevant features according to the mean absolute SHAP values for the Gradient Boosting classifiers. The left panel shows the result for the standard labelling used in \citep{di2021populist}, while the right panel shows the result for manually coded populist sentences.}
\label{fig:shap_values_IT}
\end{figure}

\begin{figure}[ht]
\centering
\includegraphics[width=0.5\linewidth]{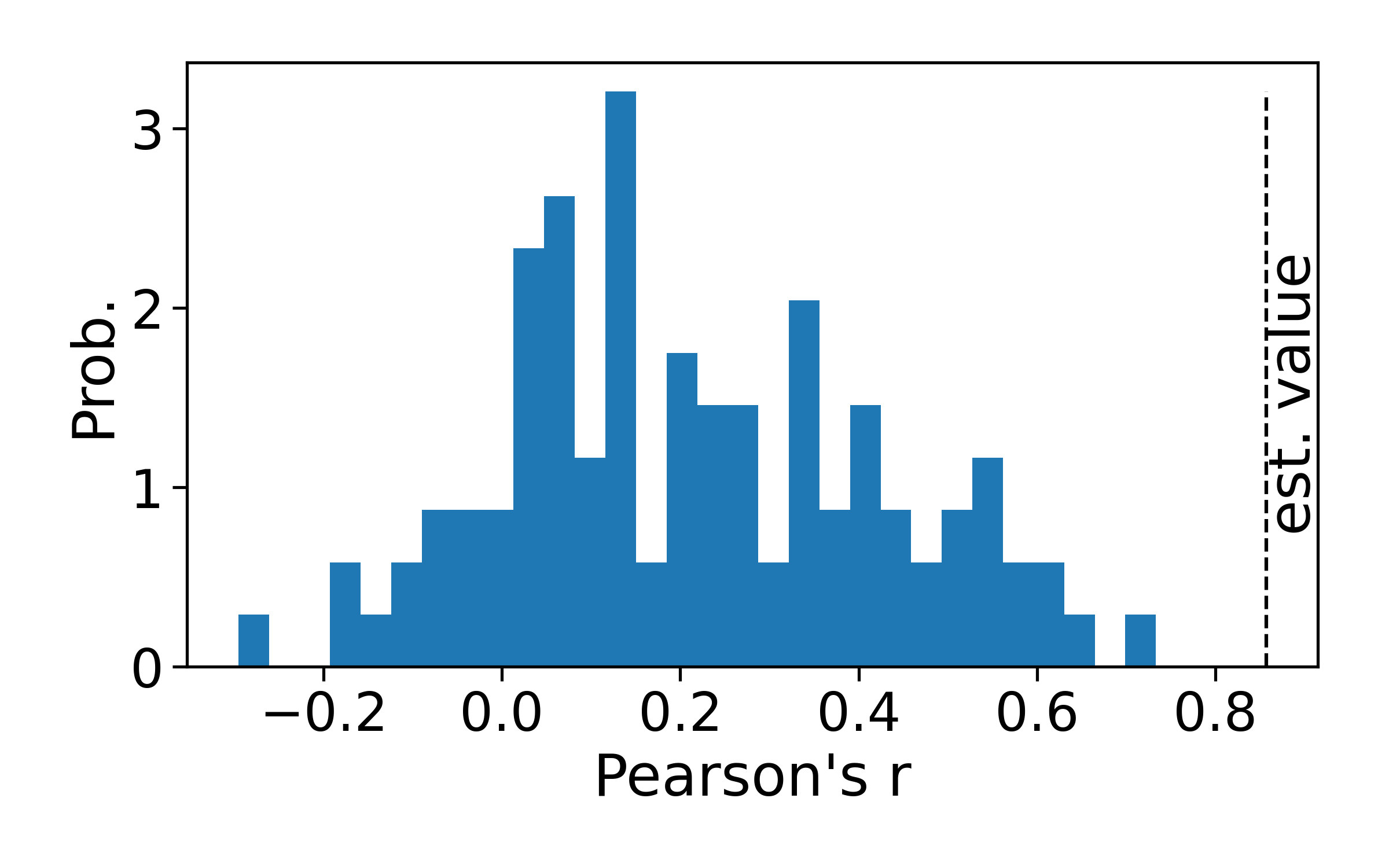}
\caption{Pearson's correlation coefficient between the Manual Coding Populist score and the standard Populist Score computed using the 100 reshuffled manifesto data. The vertical dashed line is the coefficient value for the non-reshuffled dataset also shown in Table~\ref{tab:old_valid_check}.}
\label{fig:manual_resh}
\end{figure}

Our method does not control for the manifesto content and length. It can be considered a way for comparing manifestos belonging to non-populist parties to those belonging to prototypical populist parties within the same country. In \citet{di2021populist} we compare our score for the Italian case with another built by manually labelling sentences as "populist" and "non-populist". In this case, a party's score indicates the fraction of populist sentences in its manifesto. Fig.~\ref{fig:shap_values_IT} shows the SHAP values for the Gradient Boosting classifiers trained on with the manually coded data and the party-based labelled data of \citet{di2021populist}. In Fig.~\ref{fig:shap_values_IT} we show the first 15 most relevant features according to the mean absolute SHAP values for the two cases. We see that, in principle, the two methods identify slightly different information. Among the most important features for manual coding we find \emph{stat} (state), \emph{cittadin} (citizen), \emph{pagh} (pay), \emph{europ} (Europe), etc. In the other case, we still have some relevant words but also others with a more ambiguous interpreatation (e.g. \emph{occor} that can be ``must'' or ``have to'', probably associated with the parties' communication style). It is worth noticing that in both cases, feature importance highly depends on the part of data used for training the classifiers.
The correlations found in \citet{di2021populist} with the manual coding score and shown again in Table \ref{tab:old_valid_check} suggest that our method based on a rough labelling of manifesto sentences leads to similar results. However, this could be again due to our method's bias in the score. Fig.~\ref{fig:manual_resh} shows the same Pearson's coefficient distributions between the manual coding score and the ones obtained with the 100 random reshuffles of the (party, year) labels. Again, we find that the correlations in the reshuffled case are lower than those found in the correct case.
\FloatBarrier

\section*{Acknowledgments}

We would like to acknowledge Michael Jankowski (Institute for Social Sciences, University of Bremen) and Robert A. Huber (Department of Political Science, University of Salzburg) that found the issues with the Austrian and German datasets.

\end{document}